\documentclass[sigconf,nonacm,screen=true]{acmart}

\usepackage{graphicx}
\usepackage{amsmath}
\usepackage{hyperref}
\usepackage{xcolor}
\usepackage{multirow}
\usepackage{subcaption}
\usepackage{booktabs} 

\newcommand{\CI}[3]{#1$_{#2}^{#3}$}

\begin{document}

\title{VegaChat: A Robust Framework for LLM-Based Chart Generation and Assessment}

\author{Marko Hostnik}
\affiliation{%
  \institution{JetBrains}
  \country{}
}
\email{marko.hostnik@jetbrains.com}

\author{Rauf Kurbanov}
\affiliation{%
  \institution{JetBrains}
  \country{}
}
\email{rauf.kurbanov@jetbrains.com}

\author{Yaroslav Sokolov}
\affiliation{%
  \institution{JetBrains}
  \country{}
}
\email{yaroslav.sokolov@jetbrains.com}

\author{Artem Trofimov}
\affiliation{%
  \institution{JetBrains}
  \country{}
}
\email{artem.trofimov@jetbrains.com}

\begin{abstract}

Natural-language-to-visualization (NL2VIS) systems based on large language models (LLMs) have substantially improved the accessibility of data visualization. However, their further adoption is hindered by two coupled challenges: (i) the absence of standardized evaluation metrics makes it difficult to assess progress in the field and compare different approaches; and (ii) natural language descriptions are inherently underspecified, so multiple visualizations may be valid for the same query. To address these issues, we introduce \emph{VegaChat}, a framework for generating, validating, and assessing declarative visualizations from natural language.

We propose two complementary metrics: \emph{Spec Score}, a deterministic metric that measures specification-level similarity without invoking an LLM, and \emph{Vision Score}, a library-agnostic, image-based metric that leverages a multimodal LLM to assess chart similarity and prompt compliance. 

We evaluate VegaChat on the NLV Corpus ~\cite{nlvcorpus-10.1145/3411764.3445400} and on the annotated subset of ChartLLM ~\cite{ko2023natural}. VegaChat achieves near-zero rates of invalid or empty visualizations, while Spec Score and Vision Score exhibit strong correlation with human judgments (Pearson 0.65 and 0.71, respectively), indicating that the proposed metrics support consistent, cross-library comparison.

The code and evaluation artifacts are available at \url{https://zenodo.org/records/17062309}.

\end{abstract}





\maketitle

\section{Introduction}\label{sec:introduction}

Data visualization is central to exploratory data analysis, yet programmatic creation often requires substantial expertise with libraries such as Vega-Lite~\cite{vega-lite-7539624}, Altair~\cite{altair-VanderPlas2018}, or Matplotlib~\cite{matplotlib-Hunter:2007}. Natural Language to Visualization (NL2VIS) systems seek to mitigate this barrier, but their practical deployment remains challenging. Many existing approaches generate executable code~\cite{LIDA-dibia-2023-lida, matplotagent-yang-etal-2024-matplotagent}, which introduces security risks and complicates interactive modification (UI wizards, etc.).

\begin{figure}[t]
    \centering
    \includegraphics[width=0.90\linewidth]{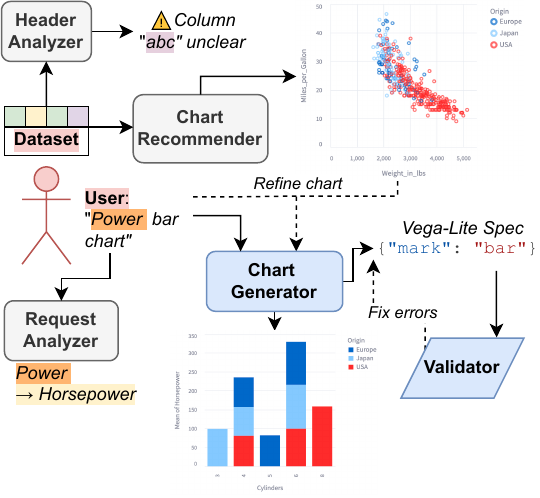}
    \caption{Schema of our chart generation system}
    \Description{A diagram showing the components of VegaChat: Header Analyzer, Chart Recommender,  Request Analyzer and the Chart Generator.}
    \label{fig:system}
\end{figure}

To address these issues, we present {\em VegaChat}, an LLM-based system that generates declarative Vega-Lite specifications. Vega-Lite (VL) provides a high-level grammar for interactive visualization with a concise JSON syntax renderable across engines~\cite{vega-2016-reactive-vega-architecture, lets-plot}. Building on this foundation, VegaChat incorporates chart recommendation and ambiguity detection to enhance real-world adoption. The declarative nature of Vega-Lite further supports secure validation and modification, enabling the development of interactive UI components—such as wizards for adjusting axes, aggregation functions, or transformations~\cite{dataformulator-wang2024dataformulator2iteratively, chartseer2022}.

Since natural language is inherently ambiguous, multiple visualizations can validly represent the same query, yet most existing metrics rely on exact matches to reference charts~\cite{datatone-10.1145/2807442.2807478, vievallm-10.2312:eurova.20241118, nvbench-10.1145/3448016.3457261}. In practice, user utterances specify what to visualize rather than how~\cite{nlvcorpus-10.1145/3411764.3445400}, so a reference chart represents only one possible outcome. To address this, we introduce two evaluation metrics: {\em Spec Score}, which measures similarity between generated and reference specifications without relying on LLMs, and {\em Vision Score}, which uses a multimodal LLM to compare chart images, making it framework-agnostic.

The main contributions of this paper are:
\begin{itemize}
    \item {\em VegaChat}, an LLM-based system for generating VL specifications from NL requests, with error handling, chart recommendation, and ambiguity detection.
    \item Two new evaluation metrics: {\em Spec Score}, which compares specifications without LLMs, and {\em Vision Score}, which uses a multimodal LLM to compare chart images across frameworks.  
    \item Comprehensive evaluation on multiple datasets demonstrating efficiency of the proposed system and metrics.  
\end{itemize}

\section{Related Work}\label{sec:related-work}

Early NL2VIS systems were rule-based~\cite{nl4dv2021, datatone-10.1145/2807442.2807478} or trained deep networks~\cite{voigt-etal-2024-plots, ncnet, rgvisnet-10.1145/3534678.3539330}. Recent work leverages pretrained LLMs to map NL to charts. Some generate an intermediate Data Visualization Query (DVQ)~\cite{nvbench-10.1145/3448016.3457261, prompt4vis-li2024prompt4vispromptinglargelanguage, nvAgent-ouyang2025nvagentautomateddatavisualization}, while others, e.g., LIDA~\cite{LIDA-dibia-2023-lida} and CoML4VIS~\cite{coml4vis-zhang-etal-2024-mlcopilot, viseval-10670425}, directly produce code in general-purpose languages~\cite{chat2vis-10121440, matplotagent-yang-etal-2024-matplotagent}.
Agent-based LLM methods~\cite{nvAgent-ouyang2025nvagentautomateddatavisualization, matplotagent-yang-etal-2024-matplotagent}, such as nvAgent, address error correction and multi-table queries~\cite{nvbench-10.1145/3448016.3457261} but incur higher latency and token costs. Other systems generate VL specs directly~\cite{VLpromptengineering-li2024visualizationgenerationlargelanguage} or via intermediate forms~\cite{ncnet, nvbench-10.1145/3448016.3457261, itergen-ugare2025itergen}. IterGen constrains output to a VL grammar subset~\cite{itergen-ugare2025itergen}. NL4DV-LLM~\cite{sah2024nl4dvllm} uses a single prompt for NL-to-data mapping, visualization recommendation, and multi-turn interaction.

\paragraph{NL2VIS Metrics}
NL2VIS metrics are either rule-based ~\cite{vievallm-10.2312:eurova.20241118, nlvcorpus-10.1145/3411764.3445400, viseval-10670425}, like our Spec Score, or LLM-based~\cite{matplotagent-yang-etal-2024-matplotagent, pandasplotbench-galimzyanov2025drawingpandasbenchmarkllms, viseval-10670425, LIDA-dibia-2023-lida}, like our Vision Score. Rule-based metrics decompose visualizations into elements (marks, encodings, transforms) and compute structured similarity scores. Unlike SVG-deconstruction metrics~\cite{viseval-10670425, VLpromptengineering-li2024visualizationgenerationlargelanguage}, which parse rendered output to extract properties, we evaluate VL specifications directly at the declarative level, combining multiple correctness aspects~\cite{vievallm-10.2312:eurova.20241118} into a single interpretable score.

The SVG-based Legality Checker~\cite{viseval-10670425, nvAgent-ouyang2025nvagentautomateddatavisualization} verifies chart type, data ordering, and axis properties through SVG parsing. Chart Readability~\cite{viseval-10670425, nvAgent-ouyang2025nvagentautomateddatavisualization} uses a multimodal LLM to assess layout, labels, and colors, but is less suitable for declarative charts, where rendering backends~\cite{vega-lite-7539624, lets-plot} determine appearance and styling varies across implementations. MatPlotBench~\cite{matplotagent-yang-etal-2024-matplotagent, pandasplotbench-galimzyanov2025drawingpandasbenchmarkllms} prompts a vision-language LLM to rate image similarity (0–100) but omits prompt compliance assessment.

Our Spec and Vision scores allow partial matches, as in NLV~\cite{nlvcorpus-10.1145/3411764.3445400}, recognizing that multiple valid visualizations may satisfy the same query. Other metrics include VIS component matching for DVQ queries~\cite{nvbench-10.1145/3448016.3457261, rgvisnet-10.1145/3534678.3539330}, which would need adaptation for VL, and others described in Section~\ref{subsec:other-evaluation-metrics}.

\section{Declarative Chart Generation}\label{sec:chart-generation}

Our system, VegaChat, follows a multi-component approach for translating NL requests into data visualizations, as shown in Figure~\ref{fig:system}.
The core of the system is a chart generation pipeline that generates Vega-Lite specifications from requests, while other components serve to improve the user experience in our prototype application.\footnote{Available at \url{https://zenodo.org/records/17062309}.}

\subsection{Chart Generator}\label{subsec:conversational-interface}

The multi-turn conversational \emph{Chart Generator} converts prompts into declarative charts, refined through follow-up interactions.
VL specifications are generated by an LLM using a few-shot prompt~\cite{VLpromptengineering-li2024visualizationgenerationlargelanguage}. 
Additionally, a NL description of the chart is generated to aid the user's understanding.

The LLM input includes the user’s request and the first five rows of the dataset in text form~\cite{pandasplotbench-galimzyanov2025drawingpandasbenchmarkllms}.
We found that GPT-4o-mini~\cite{openai2024gpt4ocard} struggles with advanced VL features (e.g., transformations, faceting). To mitigate this, we supply relevant examples in the system prompt. 
For multi-turn conversations, we instruct the model to treat follow-ups as chart refinements.

Generated specifications are validated against the schema and heuristics to catch issues such as empty but valid charts.\footnote{E.g., filtering with \texttt{"JP"} instead of \texttt{"Japan"}.}
Invalid specs are returned to the LLM in a feedback loop~\cite{matplotagent-yang-etal-2024-matplotagent, nvAgent-ouyang2025nvagentautomateddatavisualization}.

Due to the declarative form of Vega-Lite, some errors are fixed algorithmically (e.g., normalizing datetime fields from \texttt{"2006-01-01"} to \texttt{\{"year": 2006, "month": "jan", "date": 1\}}), avoiding extra LLM calls.
When schema errors cannot be fixed automatically, the LLM is tasked with correcting them, using Altair’s validation trace for additional context~\cite{altair-VanderPlas2018}.

\subsection{Auxiliary Components}\label{subsec:auxiliary-components}

To improve VegaChat’s user experience, we add auxiliary components beyond chart generation. The \emph{Header Analyzer} inspects datasets for ambiguous column names, which an LLM flags as potential obstacles for mapping requests to columns.

The \emph{Chart Recommender} proposes visualizations before any user prompt. An LLM generates diverse specifications, which are validated and auto-corrected if inconsistent with the schema. Users can then refine the suggested base plot or start from scratch.

During conversations, the \emph{Request Analyzer} checks queries for ambiguities. An LLM maps NL utterances to dataset columns, assigns confidence scores, and detects missing columns~\cite{sah2024nl4dvllm, datatone-10.1145/2807442.2807478}. If confidence falls below a threshold, users receive a warning. The analyzer also handles requests unrelated to columns or data in general. Its performance is evaluated on a small benchmark as described in Section~\ref{subsec:request-analyzer-benchmark}.

\section{Evaluation Metrics}\label{sec:metrics}

We introduce two metrics, Spec Score and Vision Score, which aim to quantify the similarity of a generated specification to a reference specification as a single number.
In both cases, we employ the flexible paradigm of aggregating multiple criteria into a single score by a weighted sum.
We also introduce a metric that accounts for empty charts.

\subsection{Specification Correctness Score}\label{subsec:spec-score}
The Specification Correctness Score (Spec Score) assesses the \textit{fuzzy} similarity between a generated specification and a reference one.
The metric aims to compare the semantic meaning of the visualizations rather than the exact match of their declarative structure or rendered visual appearance.
Emphasis is placed on mark, encoding, and transform correctness.
Stylistic aspects, such as plot titles or line colors, are ignored.
The final score is a weighted sum of F1 scores for encoding, mark, and transform correctness, with slight rewards for valid specifications to differentiate among error-free charts.  

\begin{figure}[t]
    \centering
    \includegraphics[width=\linewidth]{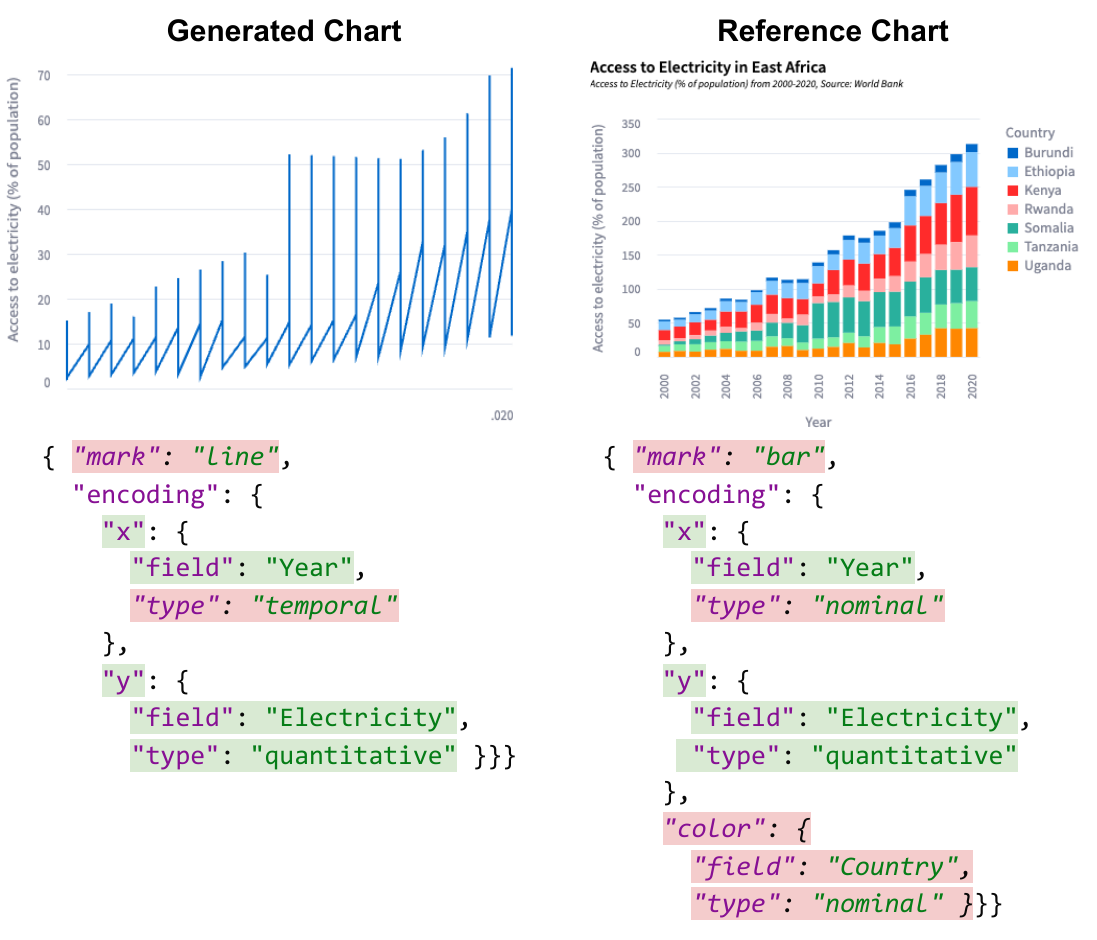}
    \caption{Example of the Spec Score metric giving a score of $50\%$}
    \Description{Comparison of a generated chart to a reference chart with Vega-Lite specifications listed. Matching fields are highlighted in green, while mismatching fields are highlighted in red.}
    \label{fig:specscore}
\end{figure}

Specifications that do not adhere to the VL schema receive a score of zero, and empty charts are heavily penalized.
Since the sum is weighted, we choose to assign a high weight mainly to the encoding correctness aspect, as it contains the most semantic information for the appearance of the final chart.
As a deterministic method, Spec Score is explainable, fast, and cost-effective, as it does not rely on LLMs for judgment. 
A visual example of the metric is shown in Figure~\ref{fig:specscore}.

Instead of using exact match comparisons~\cite{nvbench-10.1145/3448016.3457261}, we use F1 scores\footnote{We use $\beta = 2$ in the $F_\beta$ score to give more weight to recall~\cite{Rijsbergen1979} for the data encoding correctness.} for each checked VL component (e.g., encodings, mark type, transforms), inspired by work in slot filling evaluation~\cite{dstc8-google-rastogi2020towards}.
This allows us to compare lists of fields from the generated and reference charts.
It also supports cases when some fields are either missing or added in either chart.
For example, the generated chart might be missing the \texttt{color} encoding channel while still containing correct \texttt{x} and \texttt{y} channels resulting in a high precision but low recall score.

We adjust mark and encoding correctness calculations to account for partial matches.
For instance, if mark types differ (e.g., \texttt{bar} vs. \texttt{line}) but encodings are identical, we penalize the mark mismatch while keeping the overall score high, since the information conveyed remains the same.
This contrasts with prior work where mark mismatches yield zero scores~\cite{VLpromptengineering-li2024visualizationgenerationlargelanguage, viseval-10670425}.
We treat visually similar marks (e.g., circles and points) as partially equivalent (50\% penalty) and assign higher weight to mark correctness when the user's utterance explicitly mentions a mark type, detected via string matching.
For encoding correctness, we allow $x$/$y$ axis swaps~\cite{nlvcorpus-10.1145/3411764.3445400} and treat row/column faceting as equivalent.

\subsection{Vision Score}\label{subsec:vision-score}

To be able to compare our solution with other approaches that do not necessarily generate VL specifications, we introduce the Vision Score metric, which operates on images.
We use a vision-capable LLM (VLLM)~\cite{openai2024gpt4ocard} to evaluate the similarity of a generated image to a reference image~\cite{matplotagent-yang-etal-2024-matplotagent, pandasplotbench-galimzyanov2025drawingpandasbenchmarkllms, nvAgent-ouyang2025nvagentautomateddatavisualization, LIDA-dibia-2023-lida}.

The VLLM is instructed to assign scores based on visualization type, data encoding, data transformation, and plot aesthetics.
It also rates prompt compliance (i.e., whether the generated plot fulfills the user's intent) and detects empty charts.
Each dimension is scored on a discrete scale of 0, 1, or 2 points, reflecting the degree of agreement.
The dimension scores are aggregated via a weighted sum, rather than relying on a single overall score~\cite{matplotagent-yang-etal-2024-matplotagent} produced by the VLLM.
Data encoding is given the largest weight, while chart aesthetics are discounted. 

\begin{figure}[t]
    \centering
    \includegraphics[width=\linewidth]{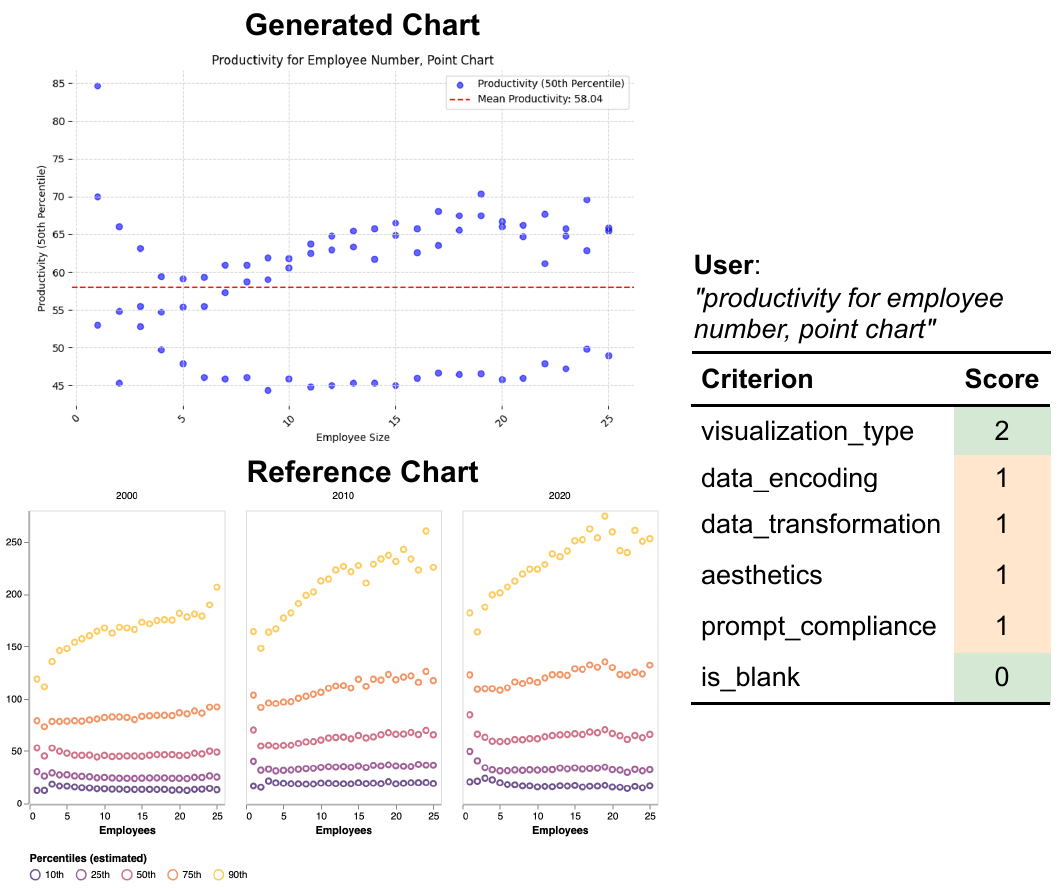}
    \caption{Example of the Vision Score metric giving a score of $58\%$}
    \Description{Comparison of a generated chart to a reference chart with the user's prompt and the resulting Vision Score criteria listed.}
    \label{fig:vision-score-benchmark}
\end{figure}

To set the individual weights of the Vision Score dimensions aggregation formula, we employ our subjective judgment.
We later verify Vision and Spec Score against diverse human judgments in Section~\ref{subsec:human-correlation} and show that the aggregation weights can be learned automatically in Section~\ref{app:human-eval-feature-importance}.
An example of the Vision Score output is shown in Figure~\ref{fig:vision-score-benchmark}.

\subsection{Spec and Vision Score Limitations}\label{subsec:spec-score-limitations}

\begin{figure}[t]
    \centering
    \includegraphics[width=0.76\linewidth]{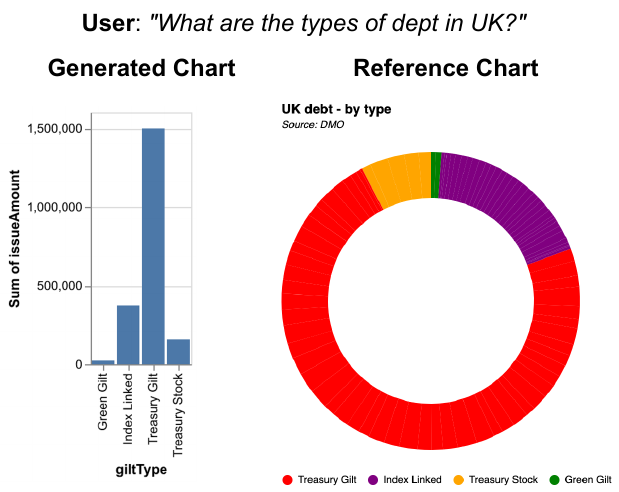}
    \caption{Example where Spec Score assigns a low score of $0.6\%$, while Vision Score assigns $78\%$}
    \Description{The generated image is a bar chart while the reference image is a donut chart.}
    \label{fig:spec_score_failure}
\end{figure}

In Figure~\ref{fig:spec_score_failure}, we identify a failure case of the Spec Score metric, which outputs a low score despite the high semantic similarity between the two charts.
The reason for the low Spec Score is that the generated and reference VL specifications are too different to reliably detect in our current implementation (e.g., the generated chart uses the \texttt{x} encoding channel, while the reference one uses \texttt{theta}).
Conversely, the Vision Score metric successfully recognizes that both charts convey the same information, penalizing only the mark type mismatch.
This example also highlights the difficulty of Question type utterances from the ChartLLM dataset (see Section~\ref{subsec:datasets}), which can be interpreted in multiple ways, leading to outputs that diverge from the reference chart.

Another downside of Spec Score is that the VL schema is complex, which requires significant attention to the details of the schema in the implementation of the metric.
For instance, many transformations can be equivalently specified inline in the encoding channels or in a separate top-level transformations array.
This makes naive, structure-based comparisons of specifications error-prone and insufficient.
Furthermore, the VL schema has undergone multiple revisions throughout its lifetime and remains in active development, implying that Spec Score might need to be adapted in the future in case of breaking changes.
In all our experiments we use the v5 version of the VL schema.

ChartLLM, and real-world specifications in general, often include \emph{interactive} charts that users can manipulate.
Such interactive features are not captured by either of our current metrics, despite being crucial elements in many practical visualizations.
Additionally, a limitation of declarative formats like Vega-Lite is that the set of possible transformations is predefined, whereas code-based approaches can flexibly perform arbitrary data transformations.

\subsection{ECR, MPB and SEVQ}\label{subsec:other-evaluation-metrics}
The Empty Chart Rate (ECR) quantifies the proportion of outputs that are either empty or otherwise result in an invalid chart due to syntax or schema validation errors.
We use ECR as a stricter version of Visualization Error Rate (VER)~\cite{LIDA-dibia-2023-lida, vievallm-10.2312:eurova.20241118}, which only checks whether the code executes successfully without errors.
For VL charts, empty outputs are algorithmically detected using Vega's scene graph representation~\cite{vega-2016-reactive-vega-architecture}.
For image-based outputs (e.g., Matplotlib charts), we use the Vision Score's empty chart detection capability.
As an alternative to Vision Score, we evaluate also the VLLM metric introduced by MatPlotBench (MPB)~\cite{matplotagent-yang-etal-2024-matplotagent, pandasplotbench-galimzyanov2025drawingpandasbenchmarkllms}.
MPB rates how well the generated chart matches the reference chart (0-100), but without taking into account the prompt that was used to generate the charts. 
We additionally use the Self-Evaluated Visualization Quality (SEVQ) metric~\cite{LIDA-dibia-2023-lida}, which relies on LLM-based judgments across six dimensions without requiring a reference chart for comparison.
The final SEVQ score is the average of the individual dimension scores.

\section{Experimental Results}\label{sec:experimental-results}

We use the GPT-4o-mini~\cite{openai2024gpt4ocard} model with the temperature set to zero for all components of VegaChat, as well as for the baselines, LIDA~\cite{LIDA-dibia-2023-lida} and CoML4VIS~\cite{viseval-10670425, coml4vis-zhang-etal-2024-mlcopilot}.
For the Vision Score, MPB and SEVQ metrics, we use the GPT-4o~\cite{openai2024gpt4ocard} model.
The metrics are reported in percentages with 95\% confidence intervals.

\subsection{Datasets}\label{subsec:datasets}


Our evaluation focuses on datasets containing VL specifications and real human 
utterances instead of synthetic ones.
We evaluate our approach using two publicly available datasets.
First, NLV Corpus~\cite{nlvcorpus-10.1145/3411764.3445400} provides 814 relatively simple utterance sets for 30 visualizations (ten visualizations $\times$ three datasets), out of which 59 examples contain sequential (multi-turn) utterances.
Second, we use the annotated section of the ChartLLM dataset~\cite{chart-llm-10.1145/3613904.3642943}, which includes 48 real-world sets, each paired with three human-written prompts (comprising commands, queries, and questions), totaling 144 examples.

We limit ourselves to single-table scenarios because NLV Corpus and ChartLLM's 
annotated subset provide VL reference specifications only for single tables, 
which are required for Spec Score computation. While nvBench~\cite{nvbench-10.1145/3448016.3457261} 
and VisEval~\cite{viseval-10670425} include multi-table cases, they lack VL 
reference specifications needed for our metrics. 
Multi-table support through integration with NL-to-SQL systems remains future work.

\begin{table*}[t]
    \centering
    \caption{Evaluation results of chart generation approaches on ChartLLM and the non-sequential part of NLV}
    \label{tab:chart-generation-comparison}
    \begin{tabular}{@{}llcccccc@{}}
        \toprule
        Dataset & Model & Spec Score & Vision Score & MPB & SEVQ & VER & ECR \\
        \midrule

        \multirow{4}{*}{NLV} &
           LIDA/Altair     & \CI{50.4}{47.6}{53.2}          & \CI{65.8}{63.3}{68.4}   &  \CI{37.6}{35.2}{40.1}        & \CI{88.6}{87.8}{89.5}          & \CI{30.6}{27.4}{33.8} & \CI{35.1}{31.7}{38.4} \\
         & LIDA/Matplotlib & /                              & \CI{74.7}{72.4}{76.9}    & \CI{52.3}{49.9}{54.6}      & \CI{81.3}{80.3}{82.3}          & \CI{8.9}{7.0}{11.0}   & \CI{10.3}{8.3}{12.6} \\
         & CoML4VIS          & /                              & \CI{81.6}{79.9}{83.4}   & \CI{58.8}{56.4}{61.2}       & \CI{86.6}{85.7}{87.4}          & \CI{0.7}{0.3}{1.5}    & \CI{2.4}{1.5}{3.7} \\
         & VegaChat (ours)   & \CI{\textbf{83.9}}{82.5}{85.2} & \CI{\textbf{85.1}}{83.6}{86.5} & \CI{\textbf{61.9}}{59.4}{64.3} & \CI{\textbf{90.6}}{89.8}{91.3} & \textbf{0.0}          & \CI{\textbf{0.3}}{0.0}{0.9} \\

        \midrule

        \multirow{4}{*}{ChartLLM} &
           LIDA/Altair     & \CI{46.2}{40.1}{52.3}          & \CI{52.0}{45.6}{58.5}   & \CI{24.3}{19.6}{28.9}       & \CI{\textbf{80.5}}{77.5}{83.6} & \CI{9.5}{5.6}{15.9}   & \CI{21.4}{15.1}{29.4} \\
         & LIDA/Matplotlib & /                              & \CI{51.3}{45.1}{57.4}   & \CI{26.7}{21.9}{31.6}      & \CI{75.0}{72.0}{78.0}          & \CI{15.9}{10.3}{23.0} & \CI{15.9}{10.3}{23.0} \\
         & CoML4VIS          & /                              & \CI{52.8}{46.7}{58.9}    & \CI{\textbf{30.5}}{25.5}{35.6}      & \CI{71.0}{67.5}{74.5}          & \CI{7.9}{4.0}{13.5}   & \CI{14.3}{8.7}{21.4} \\
         & VegaChat (ours)   & \CI{\textbf{52.6}}{47.7}{57.5} & \CI{\textbf{56.7}}{51.1}{62.4} & \CI{26.2}{21.3}{31.1} & \CI{80.1}{77.3}{82.9}          & \textbf{0.0}          & \CI{\textbf{0.8}}{0.0}{4.8} \\

        \bottomrule
    \end{tabular}
\end{table*}

\subsection{Chart Generation Results}\label{subsec:chart-generation-results}

\paragraph{Chart Generation Results}
We compare our chart generation approach, described in Section~\ref{sec:chart-generation}, to existing approaches in Table~\ref{tab:chart-generation-comparison}.
We use LIDA~\cite{LIDA-dibia-2023-lida} to generate either Matplotlib~\cite{matplotlib-Hunter:2007} or Altair~\cite{altair-VanderPlas2018} code.
Altair is a library based on Vega-Lite; consequently, its generated code can be converted to VL specifications, allowing the use of our Spec Score metric.
CoML4VIS~\cite{coml4vis-zhang-etal-2024-mlcopilot, viseval-10670425} is used to generate Matplotlib code.
In Table~\ref{tab:chart-generation-comparison}, we only use the non-sequential part of NLV, while results for its sequential part are presented in Table~\ref{tab:multi-turn-eval}.
We remove six \texttt{geoshape} mark type examples from the ChartLLM dataset, as they are not supported by our approach.
Table~\ref{tab:chart-generation-comparison} also contains SEVQ scores for comparison, although SEVQ only takes into account the generated chart and not the reference chart.
From the results we observe the efficacy of our approach in minimizing the number of empty or otherwise invalid charts (VER and ECR) compared to LIDA\@.
We attribute this to the error correction mechanism of our approach, which is further validated in Section~\ref{subsec:ablation-studies}.
We also observe that Spec and Vision scores are significantly higher for the NLV dataset, which indicates a lower difficulty level for NLV examples compared to ChartLLM.
The difficulty of ChartLLM is further examined in Section~\ref{subsec:ablation-studies}.

\paragraph{Multi-turn Conversations}

\begin{table}[!b]
    \caption{Multi-turn evaluation results on the sequential part of NLV}
    \label{tab:multi-turn-eval}

    \begin{tabular}{@{}lccc@{}}
        \toprule

        \multicolumn{4}{@{}l}{\textbf{Concatenated multi-turn prompts}} \\

        \midrule
        Model & Spec Score & Vision Score & ECR \\
        \midrule

        LIDA/Altair     & \CI{70.8}{62.8}{78.9}          & \CI{76.3}{68.4}{84.2}          & \CI{13.6}{6.8}{25.4} \\
        LIDA/Matplotlib & /                              & \CI{78.0}{70.1}{85.8}          & \CI{10.2}{3.4}{20.3} \\
        CoML4VIS          & /                              & \CI{\textbf{86.5}}{81.9}{91.0} & \textbf{0.0} \\
        VegaChat (ours)   & \CI{\textbf{86.5}}{82.3}{90.6} & \CI{85.7}{80.8}{90.6}          & \textbf{0.0} \\

        \bottomrule
    \end{tabular}

    \vspace{1ex} 

    \begin{tabular}{@{}lccc@{}}
        \toprule

        \multicolumn{4}{@{}l}{\textbf{Simulated multi-turn prompts}} \\

        \midrule
        Model & Spec Score & Vision Score & ECR \\
        \midrule

        LIDA/Altair     & \CI{49.7}{39.9}{59.4}          & \CI{54.4}{43.4}{65.3}          & \CI{32.2}{20.3}{45.8} \\
        LIDA/Matplotlib & /                              & \CI{72.1}{64.2}{80.0}          & \CI{8.5}{3.4}{18.6} \\
        CoML4VIS          & /                              & \CI{62.0}{52.7}{71.3}          & \CI{11.9}{5.1}{22.0} \\
        VegaChat (ours)   & \CI{\textbf{84.3}}{80.1}{88.4} & \CI{\textbf{82.7}}{77.3}{88.2} & \textbf{0.0} \\

        \bottomrule
    \end{tabular}

\end{table}

For practical real-world usage, NL2VIS systems should be capable of multi-turn conversations to enable the refining of generated charts~\cite{LIDA-dibia-2023-lida, nlvcorpus-10.1145/3411764.3445400, sah2024nl4dvllm}.
In Table~\ref{tab:multi-turn-eval}, we evaluate our approach in the multi-turn setting using the sequential part of the NLV dataset~\cite{nlvcorpus-10.1145/3411764.3445400}.
We simulate the multi-turn setting in two ways: either by running the system sequentially for every prompt or by concatenating all multi-turn prompts into a single new-line delimited prompt~\cite{laban2025llmslostmultiturnconversation}.
For LIDA, we use its \textit{edit} module for follow-up prompts, and for CoML4VIS, we use its \textit{fix code} module.
From the results, we observe that the concatenation approach yields charts that more closely resemble the reference charts compared to the simulated sequential approach.
A problem with the simulated approach is that the final chart output largely depends on the success of the generated chart for the first prompt, as further prompts are treated as refinements of the initial chart.
On the other hand, the concatenation approach provides the system with the full chart description as a single prompt, which makes adherence to the user's intent more likely.
We stress that concatenating all multi-turn prompts is an implementation detail that is invisible to the end user.

\begin{figure}[t]
    \centering

    \begin{subfigure}{\linewidth}
        \centering
        \includegraphics[width=\linewidth]{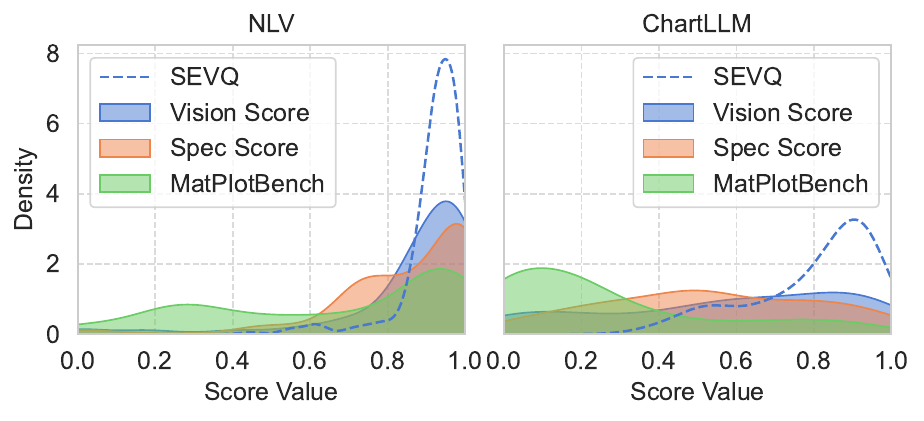}
    \end{subfigure}


    \begin{subfigure}{\linewidth}
        \centering
        \includegraphics[width=\linewidth]{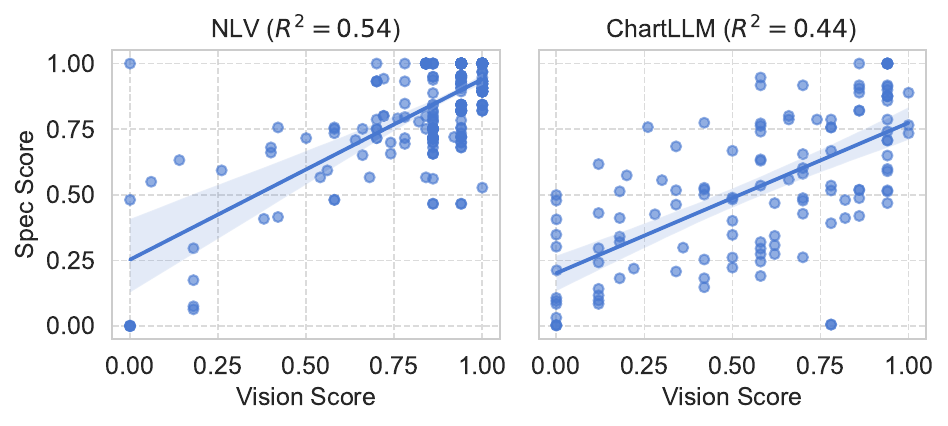}
    \end{subfigure}

    \caption{Distribution of VegaChat scores on the NLV and ChartLLM datasets}
    \Description{The top part displays distributions of SEVQ, Vision Score, Spec Score and MatPlotBench on the NLV andlChartLLM datasets. The bottom part shows scatter plots with linear regression lines, illustrating a positive correlation between Spec Score and Vision Score for the same datasets and VegaChat results.}
    \label{fig:combined_score_analysis}
\end{figure}

\paragraph{Comparison of Metrics}

The high correlation between Spec Score and Vision Score in Figure~\ref{fig:combined_score_analysis} indicates that within the domain of VL, using deterministic metrics like Spec Score can suffice, saving time and costs associated with using LLMs to judge outputs.
It also indicates that using LLMs for judging outputs can be a valid alternative to hand-crafted metrics like Spec Score.
The correlation is higher for the NLV dataset, likely due to its simpler examples, which leads to less ambiguity between the generated and reference charts.
We observe that judgments made by the Vision Score VLLM are biased towards higher scores compared to Spec Score, possibly due to the failure cases mentioned in Section~\ref{subsec:spec-score-limitations}. 
We hypothesize that the SEVQ score exhibits less variability than Spec Score and Vision Score because it does not take the reference chart into account.
Likewise we hypothesize that MatPlotBench (MPB) scores are skewed towards lower scores due to not taking the user's prompt into account when scoring.

\begin{table}[htb]
    \caption{Request Analyzer benchmark results on the NLV and ChartLLM datasets, showing the proportion of correctly classified examples}
    \label{tab:request-analyzer-benchmark}

    \begin{tabular}{@{}lcccc@{}}
        \toprule

        \multicolumn{5}{@{}l}{\textbf{NLV}} \\

        \midrule
        Sensitivity & Normal & Drop One & Drop All & Total \\
        \midrule

        low     & \textbf{100} & \CI{57.1}{48.8}{65.5} & \CI{74.4}{67.2}{81.6} & \CI{77.3}{73.2}{81.3} \\
        medium  & \CI{99.2}{97.8}{100} & \CI{72.4}{65.0}{79.8} & \CI{86.8}{81.5}{92.2} & \CI{86.2}{83.0}{89.4} \\
        high    & \CI{96.0}{92.7}{99.3} & \CI{\textbf{76.4}}{69.5}{83.3} & \CI{\textbf{87.4}}{82.2}{92.7} & \CI{\textbf{86.6}}{83.5}{89.8} \\

        \bottomrule
    \end{tabular}

    \vspace{1ex} 

    \begin{tabular}{@{}lcccc@{}}
        \toprule

        \multicolumn{5}{@{}l}{\textbf{ChartLLM}} \\

        \midrule
        Sensitivity & Normal & Drop One & Drop All & Total \\
        \midrule

        low     & \CI{\textbf{82.2}}{75.8}{88.6} & \CI{46.6}{38.1}{55.2} & \CI{68.5}{59.5}{77.6} & \CI{65.8}{61.1}{70.6} \\
        medium  & \CI{70.2}{62.5}{77.8} & \CI{66.7}{58.6}{74.7} & \CI{79.0}{71.1}{86.9} & \CI{71.4}{66.8}{76.0} \\
        high    & \CI{63.1}{55.0}{71.2} & \CI{\textbf{72.6}}{65.0}{80.2} & \CI{\textbf{83.8}}{76.6}{91.0} & \CI{\textbf{72.2}}{67.6}{76.7} \\

        \bottomrule
    \end{tabular}

\end{table}

\subsection{Request Analyzer Benchmark}\label{subsec:request-analyzer-benchmark}

To evaluate the effectiveness of the Request Analyzer described in Section~\ref{sec:chart-generation}, we create a small benchmark bootstrapped from the NLV and ChartLLM datasets.
For every example in the original dataset, we generate two additional examples by dropping specific columns from the input tabular dataset and checking if the Request Analyzer can detect the missing columns.
For the first example, we drop a \emph{single} random column that is among one of the used encoding fields in the reference VL specification.
In this case, the Request Analyzer should detect the missing dropped column.
For the second example, we drop \emph{all} columns that are used in the reference specification. 
In this case, the Request Analyzer should detect that the user's request is infeasible.
For the \emph{normal} case (i.e., no dropped columns), the Request Analyzer should trigger no warnings.\footnote{In reality, some NL requests are inherently ambiguous, meaning we can expect false positive detections.}
The results are shown in Table~\ref{tab:request-analyzer-benchmark}, grouped by the sensitivity threshold of the Request Analyzer.
Each table cell shows the proportion of examples for which the Request Analyzer either correctly detected a missing column or correctly determined that no warning was necessary.
The \emph{Total} column is calculated for the combined \emph{Normal, Drop One} and \emph{Drop All} examples.
Based on the results, we decide to use the \emph{low} sensitivity threshold to avoid false positive warnings for the \emph{normal} case.

\subsection{Ablation Studies}\label{subsec:ablation-studies}

\paragraph{Error Correction Feedback Loop}

\begin{figure}[t]
    \centering
    \includegraphics[width=0.62\linewidth]{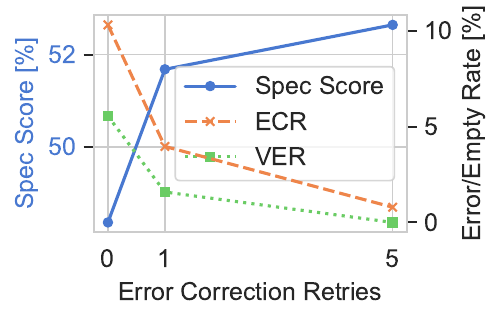}
    \caption{VegaChat results on ChartLLM based on the number of error correction retry attempts}
    \Description{Plot showing Spec Score, Empty Chart Rate and Visualization Error Rate in relation to the number of error correction retry attempts.}
    \label{fig:ec-ablation}
\end{figure}

Figure~\ref{fig:ec-ablation} shows the benefits of the error-retrying feedback loop, which instructs the LLM to fix errors (if any) in the generated VL specifications.
A significant improvement is visible in the Empty Chart Rate (ECR) metric.
In the NLV dataset (not shown), improvements are less significant, simply because there are fewer errors to correct when generating simpler charts.
The Visualization Error Rate (VER) drops to zero while ECR remains positive, highlighting the need to distinguish between valid but empty charts and error-producing ones.
Although VegaChat achieves near-zero VER and ECR scores when using five error correction attempts, as seen in Table~\ref{tab:chart-generation-comparison}, it should be noted that the other approaches we compare against do not use self-correction.

\begin{figure}[t]
    \centering
    \begin{subfigure}[t]{0.495\columnwidth}
        \centering
        \vspace{0pt}
        \includegraphics[width=\linewidth]{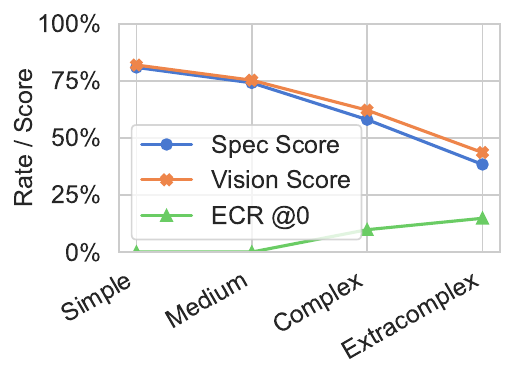}
        \label{fig:difficulty_scores_sub}
    \end{subfigure}
    \hfill
    \begin{subfigure}[t]{0.495\columnwidth}
        \centering
        \vspace{0pt}  
        \includegraphics[width=\linewidth]{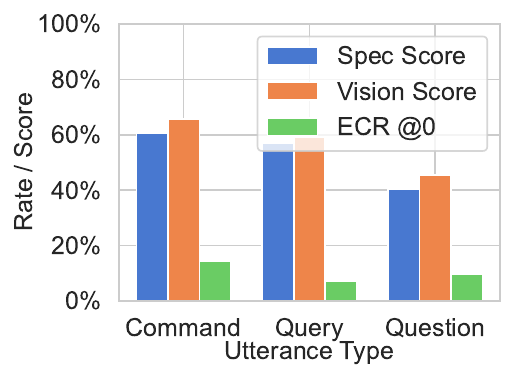}
        \label{fig:utterance_type_scores_sub}
    \end{subfigure}

    
    \caption{ChartLLM results for VegaChat, grouped by example difficulty and utterance type}
    \Description{Two side-by-side plots analyzing ChartLLM results using VegaChat. The left plot shows scores categorized by example difficulty. The right plot shows scores categorized by utterance type.}
    \label{fig:chart_llm_ablation}
\end{figure}

\paragraph{ChartLLM Difficulty and Utterance Types}
In Figure~\ref{fig:chart_llm_ablation}, we explore the performance of VegaChat on the ChartLLM dataset based on the predefined example difficulty levels and utterance types.
The Empty Chart Rate (ECR) is shown before any error correction attempts to be more informative, as the rates become negligible after five attempts as seen in Figure~\ref{fig:ec-ablation}.
Unsurprisingly, the Spec Score and Vision Score metrics decline with increasing example difficulty, while ECR increases.
For the utterance type results, we notice that Question type utterances achieve lower Spec and Vision scores compared to Command and Query utterance types.
We hypothesize that this occurs because questions leave more room for interpretation, making comparison with the reference chart problematic, especially since ChartLLM uses the same reference chart for all three utterance types.
An example of this problem is shown in Figure~\ref{fig:spec_score_failure}.
Additionally, manually inspecting the generated charts and their corresponding reference charts reveals that reference charts often contain more information than was requested by the user's prompt, which leads to Spec and Vision scores penalizing the generated chart.
Despite lower Spec and Vision scores, ECR is not problematic for Question type utterances, which indicates that the generated charts are not empty and could still be useful to the user despite their lower resemblance to the reference chart.

\subsection{Correlation with Human Judgment}\label{subsec:human-correlation}

To verify the alignment of our proposed metrics with human judgment we create a dataset composed of 171 random examples from the NLV and ChartLLM datasets, sampled using all models described in Section~\ref{subsec:chart-generation-results} to ensure our metrics are agnostic to different visualization frameworks.
Ten human annotators were tasked to rate how well the generated plots match the reference plots given the prompt used to generate the plot.
Each example was rated three times by different annotators to achieve reliable ratings.

\begin{table}[htbp]
    \centering
    \caption{Pearson's correlation coefficient to human judgments}
    \label{tab:human-eval-correlation}
    \begin{tabular}{@{}cccc@{}}
        \toprule
        Spec Score & Vision Score & MatPlotBench & SEVQ  \\
        \midrule
        \CI{0.65}{0.44}{0.76} & \CI{0.71}{0.60}{0.78} & \CI{0.73}{0.64}{0.79} & \CI{0.46}{0.30}{0.58}  \\
        \bottomrule
    \end{tabular}
\end{table}

\begin{figure}[t]
    \centering
    \includegraphics[width=0.62\linewidth]{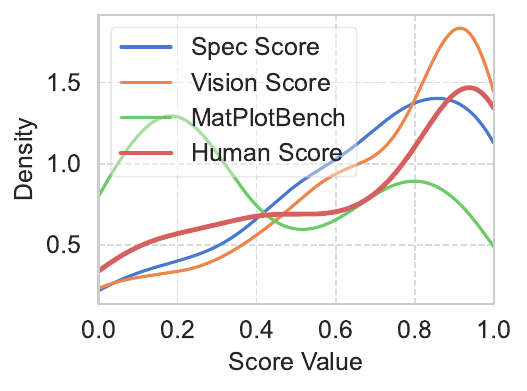}
    \caption{Metric distributions compared to human judgments}
    \Description{A comparison of the distributions of Spec Score, Vision Score and MatPlotBench to Human Score.}
    \label{fig:human-eval-distributions}
\end{figure}

We calculate the correlation between different metrics and human scores in Table~\ref{tab:human-eval-correlation}.
The results indicate that both Spec Score and Vision Score align well with human judgment, similarly to MatPlotBench.
We note that Spec Score is calculated only for Vega-Lite examples, while the rest are calculated for all examples.
In Section~\ref{app:human-eval-feature-importance} we investigate the assigned Vision Score dimensions weights and show that the correlation to human judgments can be further improved by machine learning.
In Figure~\ref{fig:human-eval-distributions}, we observe that the Vision and Spec Score distributions align well with human scores and note that the distributions are skewed toward higher values mainly due to the simple examples of NLV, as seen also in Figure~\ref{fig:combined_score_analysis}.

From the distributions in Figure~\ref{fig:human-eval-distributions} we observe that Vision Score is skewed towards higher ratings compared to MPB, which we attribute to the fact that Vision Score takes prompt compliance into account, while MPB does not.
Prompt compliance likely had an impact on human annotators, as the annotators were also shown the prompts that were used to generate the charts.
SEVQ's distribution aligns better with human scores, however, its correlation coefficient is significantly worse, as SEVQ does not take the reference chart into account.


\subsection{Vision Score Dimension Importance}\label{app:human-eval-feature-importance}

\begin{figure}[htb]
    \centering
    \includegraphics[width=0.72\linewidth]{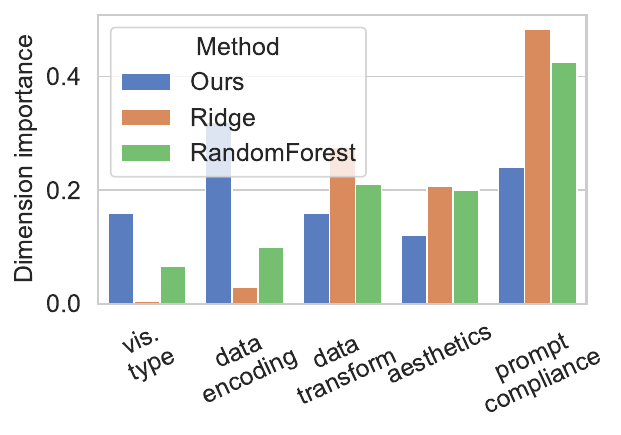}
    \caption{Vision Score dimension importance}
    \Description{A bar chart comparing the importance of each of the five Vision Score dimensions as assigned by the authors, ridge regression and random forest regression.}
    \label{fig:human-eval-feature-importance}
\end{figure}

The multi-dimensional output of Vision Score aids in the interpretability of the results and allows us to configure the final score based on our preferences about which dimensions should be weighed more than others.
Given the human evaluation judgments from Section~\ref{subsec:human-correlation}, we compare our manually assigned dimension weights to those learned by two interpretable Machine Learning (ML) models, Ridge~\cite{ridge-bhl137258} and Random Forests~\cite{random-forest-Breiman2001}.
As seen in Figure~\ref{fig:human-eval-feature-importance}, the ML models find the prompt compliance dimension to be most important, while the visualization type and data encoding dimensions are not valued as highly.
We also evaluate how well Vision Score correlates to human judgments using the learned dimension weights. 
With 10 repetitions of 5-fold cross validation, we find that using the learned weights improves Pearson's correlation coefficient by $0.02$ points over our own weights and $0.06$ points over the baseline of having equal weights for all dimensions.
These results indicate the potential of automatically tuning the outputs of Vision Score to human preferences without modifying the underlying pretrained LLM.

In an attempt to allow the Vision Score LLM to perform the final score aggregation, we prompt the model to output a single overall score, in addition to the scores for each dimension. 
The resulting overall scores are highly correlated (0.97 Pearson's correlation coefficient) to the weighted aggregation approach.
While this might be a viable approach requiring less effort, it hinders the main interpretability advantage of knowing exactly how the final score was computed from the individual dimension scores.

\section{Discussion}\label{sec:discussion}

VegaChat outperforms LIDA in both Matplotlib and Altair, with a smaller gap to CoML4VIS. For VL results, Spec Score and Vision Score are highly similar, making Spec Score a suitable substitute when VL specs exist; otherwise, Vision Score remains useful, being simpler to implement and library-agnostic.
An advantage of Vision Score compared to VLLM metrics that output a single score, such as MatPlotBench, is its interpretability, as the final score is composed of predefined dimensions, and each dimension score is accompanied by a rationale.

GPT-4o-mini still hallucinates invalid VL fields and APIs, which could be mitigated by constraining generation to a subset of the VL schema~\cite{itergen-ugare2025itergen}. LLMs also struggle with complex real-world cases like ChartLLM (Figure~\ref{fig:chart_llm_ablation}).

Multi-turn evaluation remains challenging (Table~\ref{tab:multi-turn-eval}). Simulated multi-turn prompts are unrealistic since real users adapt based on prior outputs; simulating users with an LLM may be better~\cite{laban2025llmslostmultiturnconversation}. Our concatenated approach suggests treating multi-turn interactions as single turns using the full conversation history in a single prompt.

\section{Conclusion and Future Work}\label{sec:conclusion}

We introduced \textit{VegaChat}, an LLM-based framework that turns NL queries into  Vega-Lite visualizations, together with two complementary metrics—\textit{Spec Score} and \textit{Vision Score}—that enable consistent, cross-library evaluation.

Future work includes: (i) exploring \emph{constrained/grammar-aware generation} approaches~\cite{itergen-ugare2025itergen} to ensure LLMs always emit valid Vega-Lite charts, even as the schema evolves or model pretraining cut-off dates lag; (ii) extending the metrics to capture \emph{interactive}, \emph{multi-turn}, and \emph{multi-table} visualizations so evaluation reflects real usage; and (iii) learning metric weights from human feedback to adapt Spec Score and Vision Score to specific application contexts. These directions move NL2VIS toward reliable, production-grade visual analytics.

\bibliographystyle{ACM-Reference-Format}
\bibliography{references}

@ARTICLE{vega-lite-7539624,
    author={Satyanarayan, Arvind and Moritz, Dominik and Wongsuphasawat, Kanit and Heer, Jeffrey},
    journal={IEEE Transactions on Visualization and Computer Graphics},
    title={Vega-Lite: A Grammar of Interactive Graphics},
    year={2017},
    volume={23},
    number={1},
    pages={341-350},
    doi={10.1109/TVCG.2016.2599030}}

@article{altair-VanderPlas2018,
    doi = {10.21105/joss.01057},
    url = {https://doi.org/10.21105/joss.01057},
    year = {2018},
    publisher = {The Open Journal},
    volume = {3},
    number = {32},
    pages = {1057},
    author = {Jacob VanderPlas and Brian Granger and Jeffrey Heer and Dominik Moritz and Kanit Wongsuphasawat and Arvind Satyanarayan and Eitan Lees and Ilia Timofeev and Ben Welsh and Scott Sievert},
    title = {Altair: Interactive Statistical Visualizations for Python},
    journal = {Journal of Open Source Software}
}

@Article{matplotlib-Hunter:2007,
    Author    = {Hunter, J. D.},
    Title     = {Matplotlib: A 2D graphics environment},
    Journal   = {Computing in Science \& Engineering},
    Volume    = {9},
    Number    = {3},
    Pages     = {90--95},
    publisher = {IEEE COMPUTER SOC},
    doi       = {10.1109/MCSE.2007.55},
    year      = 2007
}

@inproceedings{chart-llm-10.1145/3613904.3642943,
    author = {Ko, Hyung-Kwon and Jeon, Hyeon and Park, Gwanmo and Kim, Dae Hyun and Kim, Nam Wook and Kim, Juho and Seo, Jinwook},
    title = {Natural Language Dataset Generation Framework for Visualizations Powered by Large Language Models},
    year = {2024},
    isbn = {9798400703300},
    publisher = {Association for Computing Machinery},
    address = {New York, NY, USA},
    url = {https://doi.org/10.1145/3613904.3642943},
    doi = {10.1145/3613904.3642943},
    booktitle = {Proceedings of the 2024 CHI Conference on Human Factors in Computing Systems},
    articleno = {843},
    numpages = {22},
    location = {Honolulu, HI, USA},
    series = {CHI '24}
}

@inproceedings{vievallm-10.2312:eurova.20241118,
    booktitle = {EuroVis Workshop on Visual Analytics (EuroVA)},
    editor = {El-Assady, Mennatallah and Schulz, Hans-Jörg},
    title = {{Toward a Structured Theoretical Framework for the Evaluation of Generative AI-based Visualizations}},
    author = {Podo, Luca and Ishmal, Muhammad and Angelini, Marco},
    year = {2024},
    publisher = {The Eurographics Association},
    ISBN = {978-3-03868-253-0},
    DOI = {10.2312/eurova.20241118}
}

@inproceedings{dataformulator-wang2024dataformulator2iteratively,
author = {Wang, Chenglong and Lee, Bongshin and Drucker, Steven M. and Marshall, Dan and Gao, Jianfeng},
title = {Data Formulator 2: Iterative Creation of Data Visualizations, with AI Transforming Data Along the Way},
year = {2025},
isbn = {9798400713941},
publisher = {Association for Computing Machinery},
address = {New York, NY, USA},
url = {https://doi.org/10.1145/3706598.3713296},
doi = {10.1145/3706598.3713296},
booktitle = {Proceedings of the 2025 CHI Conference on Human Factors in Computing Systems},
articleno = {677},
numpages = {17},
location = {
},
series = {CHI '25}
}

@inproceedings{nlvcorpus-10.1145/3411764.3445400,
    author = {Srinivasan, Arjun and Nyapathy, Nikhila and Lee, Bongshin and Drucker, Steven M. and Stasko, John},
    title = {Collecting and Characterizing Natural Language Utterances for Specifying Data Visualizations},
    year = {2021},
    isbn = {9781450380966},
    publisher = {Association for Computing Machinery},
    address = {New York, NY, USA},
    url = {https://doi.org/10.1145/3411764.3445400},
    doi = {10.1145/3411764.3445400},
    booktitle = {Proceedings of the 2021 CHI Conference on Human Factors in Computing Systems},
    articleno = {464},
    numpages = {10},
    location = {Yokohama, Japan},
    series = {CHI '21}
}

@inproceedings{LIDA-dibia-2023-lida,
    title = "{LIDA}: A Tool for Automatic Generation of Grammar-Agnostic Visualizations and Infographics using Large Language Models",
    author = "Dibia, Victor",
    booktitle = "Proceedings of the 61st Annual Meeting of the Association for Computational Linguistics (Volume 3: System Demonstrations)",
    month = jul,
    year = "2023",
    address = "Toronto, Canada",
    publisher = "Association for Computational Linguistics",
    url = "https://aclanthology.org/2023.acl-demo.11/",
    doi = "10.18653/v1/2023.acl-demo.11",
    pages = "113--126"
}

@ARTICLE{viseval-10670425,
    author={Chen, Nan and Zhang, Yuge and Xu, Jiahang and Ren, Kan and Yang, Yuqing},
    journal={IEEE Transactions on Visualization and Computer Graphics},
    title={VisEval: A Benchmark for Data Visualization in the Era of Large Language Models},
    year={2025},
    volume={31},
    number={1},
    pages={1301-1311},
    doi={10.1109/TVCG.2024.3456320}}

@inproceedings{dstc8-google-rastogi2020towards,
    title={Towards scalable multi-domain conversational agents: The schema-guided dialogue dataset},
    author={Rastogi, Abhinav and Zang, Xiaoxue and Sunkara, Srinivas and Gupta, Raghav and Khaitan, Pranav},
    booktitle={Proceedings of the AAAI Conference on Artificial Intelligence},
    volume={34},
    number={05},
    pages={8689--8696},
    year={2020}
}

@article{vega-2016-reactive-vega-architecture,
    title = {Reactive Vega: A Streaming Dataflow Architecture for Declarative Interactive Visualization},
    author = {Satyanarayan, Arvind AND Russell, Ryan AND Hoffswell, Jane AND Heer, Jeffrey},
    journal = {IEEE Trans. Visualization \& Comp. Graphics (Proc. InfoVis)},
    year = {2016},
    url = {https://idl.uw.edu/papers/reactive-vega-architecture},
    doi = {10.1109/TVCG.2015.2467091}
}

@inproceedings{matplotagent-yang-etal-2024-matplotagent,
    title = "{M}at{P}lot{A}gent: Method and Evaluation for {LLM}-Based Agentic Scientific Data Visualization",
    author = "Yang, Zhiyu  and
      Zhou, Zihan  and
      Wang, Shuo  and
      Cong, Xin  and
      Han, Xu  and
      Yan, Yukun  and
      Liu, Zhenghao  and
      Tan, Zhixing  and
      Liu, Pengyuan  and
      Yu, Dong  and
      Liu, Zhiyuan  and
      Shi, Xiaodong  and
      Sun, Maosong",
    editor = "Ku, Lun-Wei  and
      Martins, Andre  and
      Srikumar, Vivek",
    booktitle = "Findings of the Association for Computational Linguistics: ACL 2024",
    month = aug,
    year = "2024",
    address = "Bangkok, Thailand",
    publisher = "Association for Computational Linguistics",
    url = "https://aclanthology.org/2024.findings-acl.701/",
    doi = "10.18653/v1/2024.findings-acl.701",
    pages = "11789--11804"
}

@inproceedings{nvbench-10.1145/3448016.3457261,
    author = {Luo, Yuyu and Tang, Nan and Li, Guoliang and Chai, Chengliang and Li, Wenbo and Qin, Xuedi},
    title = {Synthesizing Natural Language to Visualization (NL2VIS) Benchmarks from NL2SQL Benchmarks},
    year = {2021},
    isbn = {9781450383431},
    publisher = {Association for Computing Machinery},
    address = {New York, NY, USA},
    url = {https://doi.org/10.1145/3448016.3457261},
    doi = {10.1145/3448016.3457261},
    booktitle = {Proceedings of the 2021 International Conference on Management of Data},
    pages = {1235–1247},
    numpages = {13},
    location = {Virtual Event, China},
    series = {SIGMOD '21}
}

@misc{VLpromptengineering-li2024visualizationgenerationlargelanguage,
    title={Visualization Generation with Large Language Models: An Evaluation},
    author={Guozheng Li and Xinyu Wang and Gerile Aodeng and Shunyuan Zheng and Yu Zhang and Chuangxin Ou and Song Wang and Chi Harold Liu},
    year={2024},
    eprint={2401.11255},
    archivePrefix={arXiv},
    primaryClass={cs.HC},
    url={https://arxiv.org/abs/2401.11255},
}

@inproceedings{rgvisnet-10.1145/3534678.3539330,
    author = {Song, Yuanfeng and Zhao, Xuefang and Wong, Raymond Chi-Wing and Jiang, Di},
    title = {RGVisNet: A Hybrid Retrieval-Generation Neural Framework Towards Automatic Data Visualization Generation},
    year = {2022},
    isbn = {9781450393850},
    publisher = {Association for Computing Machinery},
    address = {New York, NY, USA},
    url = {https://doi.org/10.1145/3534678.3539330},
    doi = {10.1145/3534678.3539330},
    booktitle = {Proceedings of the 28th ACM SIGKDD Conference on Knowledge Discovery and Data Mining},
    pages = {1646–1655},
    numpages = {10},
    location = {Washington DC, USA},
    series = {KDD '22}
}

@INPROCEEDINGS{pandasplotbench-galimzyanov2025drawingpandasbenchmarkllms,
  author={Galimzyanov, Timur and Titov, Sergey and Golubev, Yaroslav and Bogomolov, Egor},
  booktitle={2025 IEEE/ACM 22nd International Conference on Mining Software Repositories (MSR)}, 
  title={Drawing Pandas: A Benchmark for LLMs in Generating Plotting Code}, 
  year={2025},
  volume={},
  number={},
  pages={503-507},
  doi={10.1109/MSR66628.2025.00083}}

@inproceedings{
itergen-ugare2025itergen,
    title={IterGen: Iterative Semantic-aware Structured {LLM} Generation with Backtracking},
    author={Shubham Ugare and Rohan Gumaste and Tarun Suresh and Gagandeep Singh and Sasa Misailovic},
    booktitle={The Thirteenth International Conference on Learning Representations},
    year={2025},
    url={https://openreview.net/forum?id=ac93gRzxxV}
}

@misc{nvAgent-ouyang2025nvagentautomateddatavisualization,
    title={nvAgent: Automated Data Visualization from Natural Language via Collaborative Agent Workflow},
    author={Geliang Ouyang and Jingyao Chen and Zhihe Nie and Yi Gui and Yao Wan and Hongyu Zhang and Dongping Chen},
    year={2025},
    eprint={2502.05036},
    archivePrefix={arXiv},
    primaryClass={cs.CL},
    url={https://arxiv.org/abs/2502.05036},
}

@book{Rijsbergen1979,
    author = {Rijsbergen, C. J. Van},
    edition = {2nd},
    publisher = {Butterworth-Heinemann},
    title = {Information Retrieval},
    year = 1979
}

@online{lets-plot,
    author = {JetBrains},
    title = {Lets-Plot: multiplatform plotting library built on the principles of the Grammar of Graphics},
    year = {2025},
    url = {https://lets-plot.org},
    urldate = {2025-05-16}
}

@misc{laban2025llmslostmultiturnconversation,
    title={LLMs Get Lost In Multi-Turn Conversation},
    author={Philippe Laban and Hiroaki Hayashi and Yingbo Zhou and Jennifer Neville},
    year={2025},
    eprint={2505.06120},
    archivePrefix={arXiv},
    primaryClass={cs.CL},
    url={https://arxiv.org/abs/2505.06120},
}

@misc{openai2024gpt4ocard,
    title={GPT-4o System Card},
    author={OpenAI},
    year={2024},
    eprint={2410.21276},
    archivePrefix={arXiv},
    primaryClass={cs.CL},
    url={https://arxiv.org/abs/2410.21276},
}

@inproceedings{datatone-10.1145/2807442.2807478,
    author = {Gao, Tong and Dontcheva, Mira and Adar, Eytan and Liu, Zhicheng and Karahalios, Karrie G.},
    title = {DataTone: Managing Ambiguity in Natural Language Interfaces for Data Visualization},
    year = {2015},
    isbn = {9781450337793},
    publisher = {Association for Computing Machinery},
    address = {New York, NY, USA},
    url = {https://doi.org/10.1145/2807442.2807478},
    doi = {10.1145/2807442.2807478},
    booktitle = {Proceedings of the 28th Annual ACM Symposium on User Interface Software \& Technology},
    pages = {489–500},
    numpages = {12},
    series = {UIST '15}
}

@inproceedings{voigt-etal-2024-plots,
    title = "Plots Made Quickly: An Efficient Approach for Generating Visualizations from Natural Language Queries",
    author = "Voigt, Henrik  and
      Lawonn, Kai  and
      Zarrie{\ss}, Sina",
    editor = "Calzolari, Nicoletta  and
      Kan, Min-Yen  and
      Hoste, Veronique  and
      Lenci, Alessandro  and
      Sakti, Sakriani  and
      Xue, Nianwen",
    booktitle = "Proceedings of the 2024 Joint International Conference on Computational Linguistics, Language Resources and Evaluation (LREC-COLING 2024)",
    month = may,
    year = "2024",
    address = "Torino, Italia",
    publisher = "ELRA and ICCL",
    url = "https://aclanthology.org/2024.lrec-main.1119/",
    pages = "12787--12793"
}

@ARTICLE{ncnet,
    author={Luo, Yuyu and Tang, Nan and Li, Guoliang and Tang, Jiawei and Chai, Chengliang and Qin, Xuedi},
    journal={IEEE Transactions on Visualization and Computer Graphics},
    title={Natural Language to Visualization by Neural Machine Translation},
    year={2022},
    volume={28},
    number={1},
    pages={217-226},
    doi={10.1109/TVCG.2021.3114848}}

@misc{sah2024nl4dvllm,
    title={Generating Analytic Specifications for Data Visualization from Natural Language Queries using Large Language Models},
    author={{Sah}, Subham and {Mitra}, Rishab and {Narechania}, Arpit and {Endert}, Alex and {Stasko}, John and {Dou}, Wenwen},
    year={2024},
    eprint={2408.13391},
    archivePrefix={arXiv},
    primaryClass={cs.HC},
    url={https://arxiv.org/abs/2408.13391},
    howpublished={Presented at the NLVIZ Workshop, IEEE VIS 2024}
}

@misc{prompt4vis-li2024prompt4vispromptinglargelanguage,
    title={Prompt4Vis: Prompting Large Language Models with Example Mining and Schema Filtering for Tabular Data Visualization},
    author={Shuaimin Li and Xuanang Chen and Yuanfeng Song and Yunze Song and Chen Zhang},
    year={2024},
    eprint={2402.07909},
    archivePrefix={arXiv},
    primaryClass={cs.HC},
    url={https://arxiv.org/abs/2402.07909},
}

@ARTICLE{chat2vis-10121440,
    author={Maddigan, Paula and Susnjak, Teo},
    journal={IEEE Access},
    title={Chat2VIS: Generating Data Visualizations via Natural Language Using ChatGPT, Codex and GPT-3 Large Language Models},
    year={2023},
    volume={11},
    number={},
    pages={45181-45193},
    doi={10.1109/ACCESS.2023.3274199}}

@inproceedings{coml4vis-zhang-etal-2024-mlcopilot,
    title = "{MLC}opilot: Unleashing the Power of Large Language Models in Solving Machine Learning Tasks",
    author = "Zhang, Lei  and
      Zhang, Yuge  and
      Ren, Kan  and
      Li, Dongsheng  and
      Yang, Yuqing",
    editor = "Graham, Yvette  and
      Purver, Matthew",
    booktitle = "Proceedings of the 18th Conference of the European Chapter of the Association for Computational Linguistics (Volume 1: Long Papers)",
    month = mar,
    year = "2024",
    address = "St. Julian{'}s, Malta",
    publisher = "Association for Computational Linguistics",
    url = "https://aclanthology.org/2024.eacl-long.179/",
    pages = "2931--2959"
}

@book{ridge-bhl137258,
	title = {Ridge, a computer program for calculating ridge regression estimates  },
	volume = {no.236},
	copyright = {The contributing institution believes that this item is not in copyright},
	url = {https://www.biodiversitylibrary.org/item/137258},
	publisher = {Upper Darby, Pa, Dept. of Agriculture, Forest Service, Northeastern Forest Experiment Station, 1977},
	author = {Hilt, Donald E. and Seegrist, Donald W.},
	year = {1977},
	pages = {10},
}

@inproceedings{ko2023natural,
author = {Ko, Hyung-Kwon and Jeon, Hyeon and Park, Gwanmo and Kim, Dae Hyun and Kim, Nam Wook and Kim, Juho and Seo, Jinwook},
title = {Natural Language Dataset Generation Framework for Visualizations Powered by Large Language Models},
year = {2024},
isbn = {9798400703300},
publisher = {Association for Computing Machinery},
address = {New York, NY, USA},
url = {https://doi.org/10.1145/3613904.3642943},
doi = {10.1145/3613904.3642943},
booktitle = {Proceedings of the 2024 CHI Conference on Human Factors in Computing Systems},
articleno = {843},
numpages = {22},
location = {Honolulu, HI, USA},
series = {CHI '24}
}

@ARTICLE{nl4dv2021,
    author={Narechania, Arpit and Srinivasan, Arjun and Stasko, John},
    journal={IEEE Transactions on Visualization and Computer Graphics},
    title={NL4DV: A Toolkit for Generating Analytic Specifications for Data Visualization from Natural Language Queries},
    year={2021},
    volume={27},
    number={2},
    pages={369-379},
    doi={10.1109/TVCG.2020.3030378}}

@ARTICLE{chartseer2022,
    author={Zhao, Jian and Fan, Mingming and Feng, Mi},
    journal={IEEE Transactions on Visualization and Computer Graphics},
    title={ChartSeer: Interactive Steering Exploratory Visual Analysis With Machine Intelligence},
    year={2022},
    volume={28},
    number={3},
    pages={1500-1513},
    doi={10.1109/TVCG.2020.3018724}}

@article{random-forest-Breiman2001,
  title={Random forests},
  author={Breiman, Leo},
  journal={Machine learning},
  volume={45},
  number={1},
  pages={5--32},
  year={2001},
  publisher={Springer}
}

\end{document}